\documentclass[aps,prl,twocolumn,nopacs,superscriptaddress]{revtex4}

\usepackage{amsmath}
\usepackage{amssymb}
\usepackage{bbm}
\usepackage{epsfig}
\usepackage{epstopdf}
\usepackage{color}

\usepackage{times}

\newcommand{\Tr}{{{\rm Tr}}}
\newcommand{\R}{\mathbbm{R}}
\newcommand{\Id}{\mathbf{1}}
\newcommand{\G}{\mathcal{G}}

\newtheorem{theorem}{Theorem}

\begin{document}

\title{The structure of reversible computation determines the self-duality of quantum theory}

\author{Markus P.\ M{\"u}ller}
\author{Cozmin Ududec}
\affiliation{Perimeter Institute for Theoretical Physics, 31 Caroline Street North, Waterloo, ON N2L 2Y5, Canada}
\date{February 20, 2012}

\begin{abstract}
Predictions for measurement outcomes in physical theories are usually computed by combining two distinct notions:
a \emph{state}, describing the physical system, and an \emph{observable}, describing the measurement
which is performed. In quantum theory, however, both notions are in some sense identical: outcome probabilities are given by
the overlap between two state vectors -- quantum theory is \emph{self-dual}. In this paper, we show that this notion of self-duality can be understood from a dynamical point of view.
We prove that self-duality follows from a computational primitive
called \emph{bit symmetry}: every logical bit can be mapped to any other logical bit by a reversible transformation. Specifically, we consider
probabilistic theories more general than quantum theory, and prove that every bit-symmetric theory must necessarily be self-dual.
We also show that bit symmetry yields stronger restrictions on the set of allowed bipartite states than the no-signalling principle alone, suggesting reversible time evolution
as a possible reason for limitations of non-locality.
\end{abstract}
\maketitle

A central idea of every statistical physical theory is the distinction between states and observables.
If we perform a measurement on a physical system, the state describes the preparation of the system, while the
observable corresponds to our choice of measurement. Combining the two, we obtain expectation values of measurement
outcomes.

In principle, states and observables are fundamentally distinct objects. However, in quantum theory, they turn
out to be identical: transition probabilities between two states $|\varphi\rangle$ and $|\psi\rangle$ are
given by the overlap
\begin{equation}
   {\rm Prob}(\psi\to\varphi)=|\langle\varphi|\psi\rangle|^2 = \Tr\left( |\varphi\rangle\langle\varphi| \, |\psi\rangle\langle\psi|\right).
   \label{eqBorn}
\end{equation}
More generally, the probability of obtaining an outcome described by the projector or effect operator $P$, measured on a (mixed)
quantum state $\rho$,
is given by $\Tr(\rho P)$. It is remarkable that state $\rho$ and observable $P$ are described by the same mathematical objects: up to normalization,
they are both arbitrary positive semidefinite operators~\footnote{Here we only consider finite-dimensional systems, where there is no
distinction between bounded and trace-class operators.
}. This property of \emph{self-duality}, which is most obvious in the special case~(\ref{eqBorn}),
lies at the very heart of quantum theory, and can be understood as the main ingredient in the Born rule.

In this paper we show that this remarkable property can be understood in information-theoretic terms: self-duality is
a consequence of a certain computational primitive that we call \emph{bit symmetry}. Every theory that satisfies bit symmetry -- which we
argue is necessary to allow for powerful computation -- must be self-dual.
We also prove that bit symmetry restricts the set of possible bipartite states in all theories with non-locality, including quantum theory.

{\it General probabilistic theories.}
Almost any conceivable statistical physical theory, including quantum theory and classical probability theory as special cases, can be described within the framework of general probabilistic theories~\cite{Barrett07,Barnum07,Mackey,Holevo,Hardy01}. The main physical notions are
\emph{preparations, transformations, and measurements}. Any physical system is
described by a finite-dimensional real vector space $A$. The possible preparation procedures are represented by a set of normalized
\emph{states} $\Omega_A\subset A$ (in quantum theory, $A$ is the set of self-adjoint operators on some Hilbert space,
while $\Omega_A$ is the set of density matrices). If we have
two states $\varphi,\omega\in\Omega_A$, we can think of a device which prepares either state $\varphi$ with probability $p$, or $\omega$ with
probability $1-p$, yielding the state $p\varphi+(1-p)\omega$~\cite{Holevo}. Therefore, state spaces are convex. Similarly as in quantum theory,
states will be called \emph{mixed} if they can be written as a convex combination of this form for some $0<p<1$ and $\varphi\neq\omega$, and otherwise \emph{pure}.
We also assume that state spaces are
compact, which implies that every state can be written as a finite convex combination of pure states~\cite{Barnum07}.

\begin{figure}[!hbt]
\begin{center}
\includegraphics[angle=0, width=7cm]{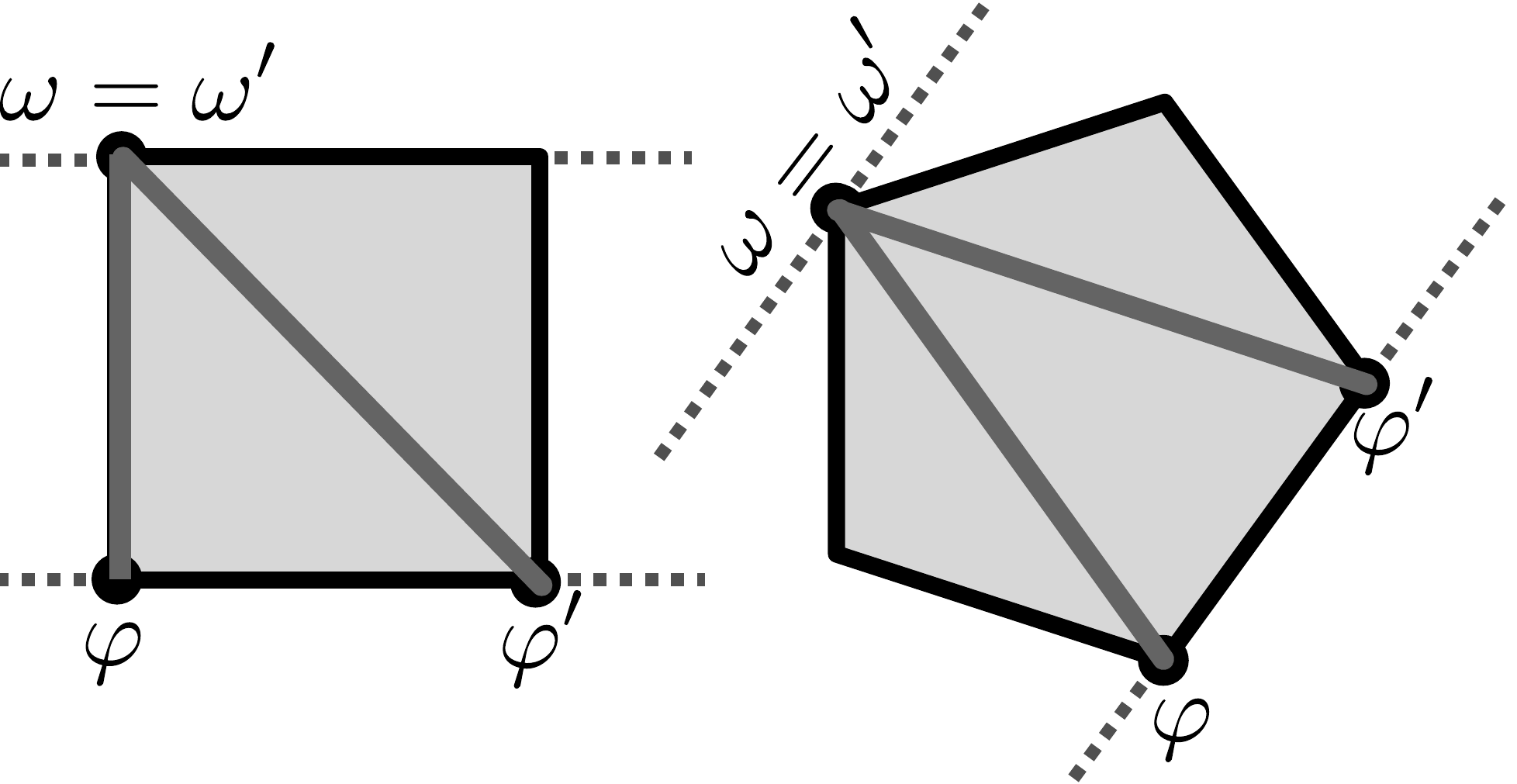}
\caption{Two state spaces: one is a square, the other a pentagon. Shown are pairs of perfectly distinguishable states
$\omega,\varphi$ and $\omega',\varphi'$. For the square, there is no symmetry which maps the pair $\omega,\varphi$ to
the pair $\omega',\varphi'$: the square state space is not bit-symmetric. For the pentagon, the pair $\omega,\varphi$ is
mapped to $\omega',\varphi'$ by a reflection across a symmetry axis. All pairs of perfectly distinguishable
pure states can be mapped to each other -- the pentagon is bit-symmetric. The dotted lines denote the level sets of
a measurement effect $E$ which distinguishes $\omega$ and $\varphi$ (and, accidentally, also $\omega'$ and $\varphi'$). That is,
the line containing $\omega$ is $\{x:E(x)=1\}$, and the line containing $\varphi$ is $\{x:E(x)=0\}$. 
For the square state space, there are two types of inequivalent logical bits: lines generated by adjacent pure states like $\omega,\varphi$,
and the square itself which is generated by diametral states like $\omega',\varphi'$.
For the pentagon -- and any other bit-symmetric theory -- 
all logical bits generated by pairs of perfectly distinguishable pure states are isometric (in this case, all pairs generate the full pentagon).}
\label{fig_bitsymmetry}
\end{center}
\end{figure}
It is important for calculations to include
unnormalized states in the framework, that is, elements of the form $\lambda\cdot\omega$ for $\lambda\geq 0$ and $\omega\in\Omega_A$. The set
of all these elements is called $A_+$. It is closed with respect to sums and convex combinations -- in convex geometry, sets of this kind are called \emph{cones}.
We assume that $A_+$ spans the whole space $A$. In quantum theory, $A_+$ is the set of positive semidefinite matrices.

In order to describe observables, consider any measurement with several possible outcomes that we perform on a state $\omega$. Denote by $E(\omega)$
the probability of obtaining one particular outcome. This must be a number between $0$ and $1$, and it must respect probabilistic
mixtures: $E\left(p\varphi+(1-p)\omega\right)=p E(\varphi)+(1-p) E(\omega)$; that is, $E$ must be linear~\cite{Barrett07}.
Linear maps $E:A\to\R$ (i.e.\ functionals) which are non-negative on all of $A_+$ will be called \emph{effects},
and the set of all effects is denoted $A_+^*$.
It is easy to see that $A_+^*$ is again a cone -- in convex geometry terms, it is called
the \emph{dual cone} of $A_+$~\cite{Aliprantis}. The normalization of states is determined by the \emph{unit} $u$, a particular element of $A_+^*$
which assigns the value one to all normalized
states: $u(\omega)=1$ for all $\omega\in\Omega_A$ (in quantum theory, we have $u(\rho)\equiv \Tr(\rho)$).
An effect $E\in A_+^*$ is called a \emph{proper effect} if $0\leq E(\omega)\leq 1$ for all states $\omega\in\Omega_A$.

In quantum theory, all effects can be written as maps $\rho\mapsto \Tr(\rho P)$, where $P\geq 0$ is a positive semidefinite matrix; it is
proper iff $P\leq\Id$. Identifying this effect with the matrix $P$, we see that $A_+^*$ can be identified with the set of positive semidefinite matrices,
such that $A_+\simeq A_+^*$. This is the notion of self-duality which will be studied in more detail in the next section. At this point, however, it is
important to note that $A_+$ and $A_+^*$ can be very different in general. As an example, consider a state space
\begin{equation}
   \Omega_A:=\left\{\left.
      (x_1,x_2,1)^T\in\R^3\,\,\right|\,\, -1\leq x_1,x_2\leq 1
   \right\}.
   \label{eqSquare}
\end{equation}
This state space looks like a square. It contains four pure states, for example $\omega=(1,1,1)^T$ and $\varphi=(-1,-1,1)^T$,
and has unit $u(x):=x_3$. Using the standard inner product on $A=\R^3$ and the pure state $\omega$, we can define a linear map $E_\omega$ by
$E_\omega(x):=\langle \omega,x\rangle=x_1+x_2+x_3$. Even though $\omega$ is a valid state, $E_\omega$
is \emph{not} a valid effect: for example $E_\omega(\varphi)=-1\not\geq 0$. For the square state space, $A_+$ and $A_+^*$
cannot be identified in this way -- they will be different no matter which inner product we use~\cite{Janotta}.\\

{\it Self-duality.} Building on the previous examples, we define a system $A$ to be \emph{self-dual}~\footnote{In the
relevant literature, this is usually called \emph{strong self-duality}, as opposed to a certain weaker form of self-duality. However, since
we do not study this weaker notion of self-duality in this paper, we drop the prefix ``strong''.
} iff
there is some inner product $\langle\cdot,\cdot\rangle$ on $A$ such that the set of effects (represented as vectors in $A$) agrees with the set of
states, $A_+^*=A_+$; that is,
\[
   A_+^*=\left\{\omega\mapsto \langle \omega,\varphi\rangle\,\,|\,\, \varphi\in A_+\right\}.
\]
Quantum theory is self-dual. To see this,
recall that for an $n$-level quantum system, the real vector space $A$ is the set of self-adjoint $n\times n$-matrices.
Consider the Hilbert Schmidt inner product on $A$, given by $\langle X,Y\rangle:=\Tr(XY)$. As we have seen above, under this inner product,
we can identify $A_+$ and $A_+^*$: both are the set of positive semidefinite matrices.

As another example, it can be shown that the square state space~(\ref{eqSquare}) is not self-dual~\cite{Janotta}, as already indicated. More generally,
regular polygons with $n$ vertices are self-dual if and only if $n$ is odd. This will become important below.\\

{\it Bit symmetry.} In addition to preparations (states) and measurements (effects), physical theories also contain a notion
of transformations. Transformations describe on the one hand possible physical time evolution, and on the other hand
possible computations that can be accomplished in the respective theory.
In this paper, we will only consider reversible transformations. This is motivated by the fact that time evolution in our universe
seems to be fundamentally reversible, and also by the conceptual analogy to the reversible circuit model in quantum computation.

Transformations must be linear (since they must respect probabilistic mixtures~\cite{Barrett07}), preserve the normalization,
and map states to states. For reversible transformations $T$, this must also be true for their inverses. Consequently, they must be
\emph{symmetries} of the state space: $T(\Omega_A)=\Omega_A$. Therefore, the set of reversible transformations on a system $A$
is a group $\G_A$, which is a subgroup of all symmetries. We assume that $\G_A$ is compact, which may be motivated on physical
grounds~\cite{Purity}. In quantum theory, $\G_A$ is the group of unitaries.

We are interested in a particular type of symmetry which connects all logical bits. To this end, we call two states $\varphi,\omega\in\Omega_A$
\emph{perfectly distinguishable} if there is a proper effect $E$ such that $E(\varphi)=0$ and $E(\omega)=1$ -- that is, if there is a
conceivable measurement device that distinguishes $\varphi$ and $\omega$ perfectly in a single run. Since all states $\psi$ have
$0\leq E(\psi)\leq 1$, the states $\varphi$ and $\omega$ must lie on opposite sides of state space:
the set of vectors $x\in A$
with $E(x)=1$ resp.\ $E(x)=0$ are two parallel supporting hyperplanes, touching the state space in $\varphi$ and $\omega$, with the full state space lying in between,
as sketched in~Fig.~\ref{fig_bitsymmetry}.

Every pair of pure and perfectly
distinguishable states $\varphi$ and $\omega$ generate a \emph{logical bit}: in terms of convex geometry, this is the \emph{face} generated
by $\varphi$ and $\omega$, that is, the smallest face~\footnote{A face of $\Omega_A$
is a convex subset $B\subseteq \Omega_A$ such that $(x,y\in \Omega_A,\lambda>0,\lambda x+(1-\lambda)y\in B)\Rightarrow x,y\in B$.
} of $\Omega_A$ containing both $\varphi$ and $\omega$.
In quantum theory, two pure states $|\varphi\rangle\langle\varphi|$ and $|\omega\rangle\langle\omega|$ are perfectly distinguishable
if and only if $\langle\varphi|\omega\rangle=0$. The logical bit that they generate is not simply the line segment making up their convex hull,
but
contains all pure states of the form $\alpha|\varphi\rangle
+\beta|\omega\rangle$ and their convex mixtures -- that is, a full Bloch ball~\footnote{Technically, when we talk about a logical bit, we also assume
a fixed choice of perfectly distinguishable pure states $\varphi,\omega$ in that face, analogous to a choice of ``basis states'' in quantum theory.}.

Now we are ready to define our main notion: a system $A$ is called \emph{bit-symmetric}, if one of the two following equivalent conditions holds:
\begin{itemize}
\item If $\varphi,\omega$ are perfectly distinguishable pure states, and so are $\varphi',\omega'$, then there is a reversible transformation
$T\in\G_A$ such that $T\varphi=\varphi'$ and $T\omega=\omega'$.
\item Every logical bit can be mapped to every other logical bit by some reversible transformation.
\end{itemize}

Quantum theory is obviously bit-symmetric: every pair of orthogonal pure states can be mapped to every other by some unitary.
It is even more symmetric than this: analogous statements hold for triples, quadruples, etc., of orthogonal pure states.
As a less trivial example, consider state spaces that are regular polygons with $n$ vertices. It turns out that these state spaces
are bit-symmetric if and only if $n$ is odd. In Fig.~\ref{fig_bitsymmetry}, this is illustrated for $n=4$ and $n=5$, i.e.\ for the square and the
pentagon.

Classical probability theory is bit-symmetric as well: the $n$-outcome state space is the set of probability distributions $(p_1,\ldots,p_n)$,
$\sum_i p_i=1$, $p_i\geq 0$. Geometrically, this convex set is a simplex, and the pure states are
of the form $(0,\ldots, 0, 1,0,\ldots,0)$ (full weight on one outcome). The reversible transformations are the permutations of the $n$ entries, which
can map every pair of pure states to every other. In fact, these ``transpositions'' generate the full group of permutations.

As the last example illustrates, bit symmetry is an important and basic computational primitive. In the context of quantum computation,
it implies that any ``entangled'' logical
bit that appears in a computation on many qubits can in principle be mapped to the first qubit (awaiting a final measurement)
without destroying coherence. In general theories, bit symmetry means that yes-no-questions which can be
answered perfectly by (irreversible) measurements may in principle also be asked ``coherently'' and be part of
a larger reversible computation.
In physical terms, it means that the state of any natural two-level system can be transferred to any
other two-level system by a suitable reversible interaction. One may argue that the time evolution of the universe would be severely
constrained if this property did not hold.\\

{\it Main Result.}  Now we prove our main theorem:
\begin{theorem}
If a state space is bit-symmetric, then it is also self-dual.

Moreover, the corresponding inner product
can be chosen to be non-negative on all states,
invariant under all reversible transformations, and to satisfy
$\langle \omega,\omega\rangle=1$ for all pure states $\omega$ and
$\langle \omega,\varphi\rangle=0$ if $\omega$ and $\varphi$ are perfectly distinguishable.
\end{theorem}
\emph{Remark.} In quantum theory, $\langle \cdot,\cdot\rangle$ is the Hilbert-Schmidt inner product between
self-adjoint matrices: $\langle X,Y\rangle=\Tr(XY)$; invariance means that $\langle UXU^\dagger, UYU^\dagger\rangle=\langle X,Y\rangle$
for all unitaries $U$. In all bit-symmetric theories, if one of $\omega$ and $\varphi$ is pure, then
$\langle \omega,\varphi\rangle=0$ implies that $\omega$ and $\varphi$ are perfectly distinguishable. However, we were
not able to prove that the same holds true in general if both are mixed.

\emph{Proof.} If $\omega\in\Omega_A$ is any pure
state, then there is always another pure state $\varphi$ that is perfectly distinguishable from $\omega$ (unless
the state space contains only a single point). Thus, bit symmetry implies \emph{transitivity}: to every pair
of pure states $\omega,\psi$, there is a reversible transformation $T\in\G_A$ such that $T\omega=\psi$. This
allows us to define a maximally mixed state $\mu^A$ as
$\mu^A:=\int_{T\in\G_A} T\omega\, dT$,
where $\omega\in\Omega_A$ is any pure state. Due to transitivity, $\mu^A$ does not depend on the choice of
$\omega$. To every state $\omega$, define its \emph{Bloch vector} $\hat\omega:=\omega-\mu^A$. Then we
can decompose the space $A$ into $A=\hat A \oplus \R\cdot\mu^A$, where $\hat A$ is the set of all points $x\in A$
with $u(x)=0$, with $u$ the unit on $A$. If $\omega$ is a state, then its Bloch vector $\hat\omega$ is an element of $\hat A$.

Since reversible transformations preserve normalization, they leave the subspace $\hat A$ invariant. According to group representation
theory~\cite{Simon}, there is an inner product $(\cdot,\cdot)$ on $\hat A$ such that $(Tx,Ty)=(x,y)$ for all $T\in\G_A$ and $x,y\in\hat A$. We may scale
this product by an arbitrary positive factor such that $(\hat \omega,\hat \omega)=1$ for all pure states $\omega$ (they all have
the same inner product due to transitivity).

Define $c:=\min_{\omega,\varphi\in\Omega_A} (\hat\omega,\hat\varphi)\leq(\hat\mu^A,\hat\mu^A)=0$ to be the minimal inner product between the Bloch vectors of any two states.
Our next step is to prove the following statements:
\begin{itemize}
\item[(i)] For all $\omega$ and $\varphi$, we have $c\leq (\hat\omega,\hat\varphi)\leq 1$, where $c<0$.
\item[(ii)] If $\omega$ is pure and $\varphi$ is arbitrary and $(\hat\omega,\hat\varphi)=c$, then $\omega$ and $\varphi$ are perfectly distinguishable.
\item[(iii)] If $\omega$ and $\varphi$ are arbitrary perfectly distinguishable states, then $(\hat\omega,\hat\varphi)=c$.
\end{itemize}
To this end, define a linear map $E_\omega:A\to\R$ for every pure state $\omega$ by linear extension of
\[
   E_\omega(\varphi):=\frac{(\hat\omega,\hat\varphi)-c}{1-c} \qquad (\varphi\in\Omega_A).
\]
Since $c\leq 0$, this is well-defined, and since $(\hat\omega,\hat\varphi)\geq c$, we have $E_\omega(\varphi)\geq 0$ for all $\varphi\in\Omega_A$.
Due to convexity of the norm $\|\hat\omega\|\equiv\sqrt{ \langle\hat\omega,\hat\omega\rangle}$, all mixed states $\omega$ satisfy $\|\hat\omega\|\leq 1$, with
equality for the pure states. Thus,
the Cauchy-Schwarz inequality implies $(\hat\omega,\hat\varphi)\leq\|\hat\omega\|\cdot\|\hat\varphi\|\leq 1$, hence $E_\omega(\varphi)\leq 1$
for all $\varphi\in\Omega_A$. In other words, for every pure state $\omega$, the map $E_\omega$ is a proper effect. Now suppose that $\omega\in\Omega_A$
is pure and $\varphi\in\Omega_A$ is arbitrary, and $(\hat\omega,\hat\varphi)=c$. Then $E_\omega(\varphi)=0$ and $E_\omega(\omega)=1$, hence $\varphi$
and $\omega$ are perfectly distinguishable. This proves (ii). Moreover, if $c=0$, we would have $(\hat\omega,\hat\mu^A)=0=c$, and so $\omega$ and $\mu^A$
would be perfectly distinguishable, which is impossible. Hence $c<0$, proving (i).

Choose $\omega,\varphi\in\Omega_A$ such that $(\hat\omega,\hat\varphi)=c$. We
can decompose $\omega$ and $\varphi$ into pure states $\omega_i$ and $\varphi_j$: $\omega=\sum_i\alpha_i\omega_i$, $\varphi=\sum_j\beta_j\varphi_j$
with $\alpha_i,\beta_j>0$. Since $c=\sum_{ij}\alpha_i\beta_j (\hat \omega_i,\hat\varphi_j)$, and $c$ is the minimal possible value, every addend must
have this value due to convexity, so $(\hat\omega_i,\hat\varphi_j)=c$ for all $i,j$. Thus $\omega_i$ and $\varphi_j$ are pure and perfectly distinguishable.
Fix some $i,j$.
Now if $\omega'$ and $\varphi'$ are another pair of pure and perfectly distinguishable states, there is a reversible transformation $T$ such that
$T\omega_i=\omega'$ and $T\varphi_j=\varphi'$, hence $(\hat\omega',\hat\varphi')=(T\hat\omega_i,T\hat\varphi_j)=(\hat\omega_i,\hat\varphi_j)=c$.
That is, every pair of perfectly distinguishable pure states has inner product $c$ between its Bloch vectors. Now suppose that $\omega$ and $\varphi$
are arbitrary perfectly distinguishable states. Decomposing them as above, it follows that every $\omega_i$ is perfectly distinguishable from every $\varphi_j$,
hence $(\hat\omega,\hat\varphi)=\sum_{ij} \alpha_i \beta_j (\hat\omega_i,\hat\varphi_j)=c$. This proves statement (iii).

Let $E$ be any effect such that $\R_0^+\cdot E$ is an exposed ray of $A_+^*$. That is, there is some $x\in A$ with the following property:
\begin{equation}
   F\in A_+^*, F(x)=0 \Rightarrow F=\lambda E\mbox{ for some }\lambda\geq 0.
   \label{eqExposing}
\end{equation}
The point $x$ defines a supporting hyperplane of $A_+^*$, touching it in the ray generated by $E$. Thus, either $F(x)\geq 0$ for all
$F\in A_+^*$, or $F(x)\leq 0$ for all $F\in A_+^*$. In the last case, we can redefine $x\mapsto (-x)$, such that $F(x)\geq 0$ for all
$F\in A_+^*$, or, in other words, $x\in (A_+^*)^*=A_+$. Since $x\neq 0$, we have $u(x)\neq 0$, and $\omega:=x/u(x)$ defines
a state $\omega\in\Omega_A$ which depends on $E$, and will be mixed in general.

Set $\lambda:=\max_{\varphi\in\Omega_A} E(\varphi)>0$ and $F:=E/\lambda$, then $F(\omega)=0$, and the set of states $\varphi$
with $F(\varphi)=1$ is a non-empty face of $\Omega_A$. Let $\omega'$ be any extremal point of that face, then it is a pure
state which is by construction perfectly distinguishable from $\omega$. Hence $(\hat\omega,\hat\omega')=c$, and so
$E_{\omega'}(x)=u(x)E_{\omega'}(\omega)=0$.
Due to~(\ref{eqExposing}), it follows that there is some $\lambda\geq 0$
such that $E_{\omega'}=\lambda E$. We have thus shown that every ray-exposed effect is of the form $\lambda' E_{\omega'}$ for some
$\lambda'>0$ and pure state $\omega'$. According to Straszewicz' Theorem~\cite{Webster}, the exposed rays are dense in the set of
extremal rays, hence every ray-extremal effect is of this form.

Now we extend $(\cdot,\cdot)$ to an inner product $\langle\cdot,\cdot\rangle$ on all of $A$. If $x,y\in A$,
use the decomposition $x=x_0\mu^A+\hat x$ with $\hat x\in\hat A$ (and similarly for $y$) and define
$\langle x,y \rangle:=\lambda x_0 y_0 + (1-\lambda)(\hat x,\hat y)$,
where $\lambda:=-c/(1-c)\in (0,1)$, since $c<0$. It is easy to check that this is an inner product,
satisfying all statements of the theorem. We can now identify linear functionals
$L:A\to\R$ with vectors $\vec L\in A$ via $L(x)=\langle \vec L,x\rangle$. Every ray-extremal effect is of
the form $E_\omega(\varphi)=\langle \omega,\varphi\rangle$
for some pure state $\omega$, hence $\vec E_\omega= \omega$. Thus, in this identification,
all extremal rays of $A_+^*$ are contained in $A_+$. Since they generate the full cone $A_+^*$, we have
$A_+^*\subseteq A_+$. On the other hand, consider an extremal ray of $A_+$; it is spanned by
some pure state $\omega$. By construction, $\langle\omega,\varphi\rangle\geq 0$ for all
$\omega,\varphi\in\Omega_A$, hence
the corresponding effect $E_\omega$ is contained in $A_+^*$.
Thus, $A_+\subseteq A_+^*$. In summary, we get $A_+=A_+^*$ under the inner product $\langle\cdot,\cdot\rangle$ -- that is, $A$
is self-dual.
\hfill $\square$

In low dimensions, bit-symmetric state spaces are rare. Using the classification of transitive state spaces in~\cite{Kimura}, it follows
that the only bit-symmetric $2$-dimensional state spaces are the unit disc and the regular polygons with an odd number of vertices.
In $3$ dimensions, there is only the unit ball (representing a qubit) and the unique regular self-dual polytope, the tetrahedron (representing a classical 4-level system).
For a different set of postulates leading to self-duality, see~\cite{Wilce}.\\

{\it Non-locality.} Given two state spaces $A$ and $B$, we can consider the set of all joint states (that is, correlations) on $AB$ which are consistent with
the no-signalling principle~\cite{Barnum07}; this is called the \emph{maximal tensor product} $A\otimes_{\max} B$ of $A$ and $B$. Explicitly,
$\Omega_{AB}$ is the set of all $\omega\in A\otimes B$ with $u^A\otimes u^B(\omega)=1$ and $E^A\otimes E^B(\omega)\geq 0$ for all
$E^A\in A_+^*, E^B\in B_+^*$.

If $A$ and $B$ are the square state space~(\ref{eqSquare}), then $A\otimes_{\max} B$ is called the ``no-signalling polytope''. It
contains so-called \emph{PR boxes}~\cite{Barrett07}
which violate the Bell-CHSH inequality by more than any quantum state. It has been asked why quantum theory does not allow for such
``maximally non-local'' states. The following theorem generalizes the results in~\cite{boxworld}:

\begin{theorem}
The maximal tensor product $A\otimes_{\max}B$ of two state spaces can only be bit-symmetric if it does not
contain any entangled states at all.
\end{theorem}
{\it Proof.} From the definition of $\Omega_{AB}$, it follows that all extremal rays of the effect cone $(AB)_+^*$ are of the form $E^A\otimes E^B$.
If $\Omega_{AB}$ is bit-symmetric, then it is self-dual; hence, all pure states (generating the state cone) are product states. Since all states are mixtures
of those, they must be unentangled.
\hfill $\square$

If $A$ and $B$ are classical $n_A$- and $n_B$-level systems, then $A\otimes_{\max} B$ is a classical $n_A n_B$-level system. It is bit-symmetric,
but does not contain any entangled states. On the other hand, any
bit-symmetric composition $AB$ of two state spaces $A$ and $B$ which \emph{does} contain entangled states (such as quantum theory)
must be a proper \emph{subset} of $A\otimes_{\max} B$ -- there are at least
some maximally non-local states of $A\otimes_{\max} B$ which $AB$ cannot contain.

While the omission of some states of $A\otimes_{\max} B$ does not in itself necessarily reduce the amount of non-locality in a theory~\cite{LocalQM},
this result still gives a hint that bit symmetry might introduce constraints on the amount of Bell inequality violations. This conjecture is further substantiated
by the findings in~\cite{Janotta}, where it was shown that a class of composites of regular $n$-gons as in Fig.~\ref{fig_bitsymmetry} satisfies the Tsirelson bound
if and only if $n$ is odd, i.e.\ the theory is locally bit-symmetric.\\

{\it Conclusions.} We have shown that self-duality, one of the defining features of quantum theory~\cite{Wilce}, follows from the computational
primitive of bit symmetry. Thus, the power of reversible computation (or, equivalently, time evolution) severely
constrains the statistical behaviour of any physical theory. We have also proven that bit symmetry restricts the 
set of allowed bipartite states, leaving the interesting open problem to quantify the consequences for
violations of Bell inequalities.
\\

{\it Acknowledgments.}
We would like to thank Christian Gogolin, Peter Janotta, Llu\'is Masanes, Jonathan Oppenheim, Tony Short, and Stephanie Wehner for discussions.
Research at Perimeter Institute is supported by the Government of Canada through Industry Canada and by the Province of Ontario through the Ministry of Research and Innovation.


\begin{thebibliography}{99}

\bibitem{Barrett07}
J.\ Barrett, \emph{Information processing in generalized probabilistic theories}, Phys.\ Rev.\ A \textbf{75} No.\ 3, 032304 (2007).

\bibitem{Barnum07}
H.\ Barnum, J.\ Barrett, M.\ Leifer, and A.\ Wilce, \emph{Generalized No-Broadcasting Theorem}, Phys.\ Rev.\ Lett.\ \textbf{99}, 240501 (2007).

\bibitem{Mackey}
G.\ Mackey, \emph{Mathematical Foundations of Quantum Mechanics} (Addison-Wesley, 1963).

\bibitem{Holevo}
A.\ S.\ Holevo, \emph{Probabilistic and Statistical Aspects of Quantum Theory} (North-Holland, New York, 1980).

\bibitem{Hardy01}
L.\ Hardy, \emph{Quantum Theory From Five Reasonable Axioms}, arXiv:quant-ph/0101012.

\bibitem{Aliprantis}
C.\ D.\ Aliprantis, R.\ Tourky, \emph{Cones and Duality} (American Mathematical Society, 2007).

\bibitem{Janotta}
P.\ Janotta, C.\ Gogolin, J.\ Barrett, and N.\ Brunner, \emph{Limits on nonlocal correlations from the
structure of the local state space}, New J.\ Phys.\ \textbf{13}, 063024 (2011).

\bibitem{Purity}
M.\ P.\ M\"uller, O.\ C.\ O.\ Dahlsten, and V.\ Vedral, \emph{Unifying typical entanglement and coin
tossing: on randomization in probabilistic theories}, arXiv:1107.6029.

\bibitem{Simon}
B.\ Simon, \emph{Representations of Finite and Compact Groups} (American Mathematical Society, 1996).

\bibitem{Webster}
R.\ Webster, \emph{Convexity} (Oxford University Press, 1994).

\bibitem{Kimura}
G.\ Kimura and K.\ Nuida, \emph{On affine maps on non-compact convex sets and some characterizations
of finite-dimensional solid ellipsoids}, arXiv:1012.5350.

\bibitem{boxworld}
D.\ Gross, M.\ M\"uller, R.\ Colbeck, and O.\ C.\ O.\ Dahlsten, \emph{All reversible dynamics in maximally
non-local theories are trivial}, Phys.\ Rev.\ Lett.\ \textbf{104}, 080402 (2010).

\bibitem{Wilce}
A.\ Wilce, \emph{Four and a Half Axioms for Finite Dimensional Quantum Mechanics}, arXiv:0912.5530.

\bibitem{LocalQM}
H.\ Barnum, S.\ Beigi, S.\ Boixo, M.\ B.\ Elliott, and S.\ Wehner, \emph{Local Quantum Measurement and
No-Signaling Imply Quantum Correlations}, Phys.\ Rev.\ Lett.\ \textbf{104}, 140401 (2010).
		
\end{thebibliography}
\end{document}